\documentstyle[aps,twocolumn,floats,prl]{revtex}
\draft
\begin{document}

\title{Nearly critical ground state of LaCuO$_{2.5}$}

\author{Matthias Troyer$^a$, M. E. Zhitomirsky$^{a,b}$,
 and Kazuo Ueda$^a$}

\address{$^a$Institute for Solid State Physics, 
University of Tokyo, Roppongi 7-22-1, Tokyo 106, Japan \\
$^b$L. D. Landau Institute for Theoretical Physics, Moscow 117334, Russia}

\maketitle
\begin{abstract}
Using a combination of analytical techniques and Quantum Monte Carlo
simulations we investigate the coupled spin ladder system
LaCuO$_{2.5}$. At a critical ratio of the interladder to intraladder
coupling $(J'/J)_c\approx0.11$ we find a quantum phase transition
between a N\'eel ordered and a disordered state. At criticality the
uniform susceptibility behaves as $\chi(T)=aT^2$ with a universal
prefactor. At intermediate temperatures the system crosses over to a
``decoupled ladders regime'' with pseudo-gap type behavior, similar to
uncoupled ladders. This can explain the gap-like experimental data for
the magnetic susceptibility of LaCuO$_{2.5}$ despite the presence of
long range N\'eel order.
\end{abstract}


\narrowtext

The unusual normal state magnetic properties of doped high-$T_c$
cuprates have led to enhanced interest in zero temperature
order-disorder transitions of quantum magnets. In particular, detailed
predictions have been made about the behavior of a two-dimensional
(2D) Heisenberg antiferromagnet by mapping it to the nonlinear sigma
model \cite{sigma-m}. They are in good agreement with experimental
measurements on La$_2$CuO$_4$. In addition to various mechanisms
proposed for 2D spin systems, long range N\'eel order at $T=0$ can
also be destroyed if a 3D antiferromagnet approaches the 1D limit due
to spatially anisotropic exchange.  Then, quantum critical behavior
and a disordered spin-liquid phase should be observed in three spatial
dimensions.

Recently, a suitable system for such type of behavior, LaCuO$_{2.5}$,
has been synthesized \cite{Hiroi}. The copper atoms in this compound
form an array of coupled spin-1/2 two-chain ladders. Isolated spin
ladders have a spin-liquid ground state and show signs of
superconducting pairing with a $d$-wave order parameter upon
doping\cite{review}. However, a marked transition to a metallic phase
takes place in LaCuO$_{2.5}$ under Sr doping, but no sign of
superconducting pairing was observed down to 5~K \cite{Hiroi}.  In
contrast superconductivity was recently found in the ladder compound
${\rm Sr}_{0.4}\rm{Ca}_{13.6}\rm{Cu}_{24}{\rm O}_{41.84}$
\cite{uehara}, which has weak and frustrated interladder couplings.
This observation makes it quite important to study in
detail the influence of the interladder coupling on the magnetic
properties of the undoped insulating phase.

First susceptibility measurements on LaCuO$_{2.5}$ were interpreted as
showing a spin-gap in the excitation spectrum \cite{Hiroi}. Subsequent
NMR and $\mu$SR studies indicated, in contrast, antiferromagnetic
ordering below $T_N\sim 110$~K \cite{Matsumoto,mu}. Normand and Rice
\cite{Normand} suggested that the magnetic state could be close 
to a transition to spin-liquid phase. In this letter we expand on this
idea and show that the apparently conflicting experimental results can
be reconciled.

The basic model for understanding these properties of LaCuO$_{2.5}$ is  
a spin-1/2 Heisenberg Hamiltonian for coupled ladders\cite{Normand}
\begin{equation}
\hat{\cal H} = J\sum_{\langle i,j\rangle}
{\bf S}_i\cdot{\bf S}_j +
J'\sum_{\langle i,j\rangle}
{\bf S}_i\cdot{\bf S}_j \ , \label{H}  
\end{equation}
which are shown schematically in Fig.~\ref{fig:lattice}. We assume for
simplicity equal rung and leg exchange constants $J$ in each ladder
and different exchange $J'$ between ladders. Notice that the
crystalline structure of LaCuO$_{2.5}$ is more complicated, having
four spins per unit cell \cite{Hiroi}.  However,
we may choose a simpler,
topologically equivalent lattice structure having only two spins per
unit cell. For $J'\approx J$ the spin system is three-dimensional and
has N\'eel order at low temperatures because the interladder coupling
does not introduce frustration. Quantum fluctuations become more and
more significant as one approaches the quasi 1D limit. Since the 1D
phase is a spin liquid with a finite gap, the magnetic order is
destroyed at some finite $J'$.

We examine the following points: (i) the critical ratio of $(J'/J)_c$
for the order-disorder transition, (ii) the low-$T$ behavior of the
uniform susceptibility $\chi$ at the critical point, and (iii)
$\chi(T)$ in the whole temperature range and for arbitrary $J'/J$.
For this we employ a combination of analytical and numerical
techniques. With the help of the renormalized spin-wave theory
\cite{Oguchi} and the bond-operator method \cite{Sachdev} we obtain
lower and upper bounds for the transition point:
$0.05<(J'/J)_c<0.12$. The quantum critical behavior of the uniform
susceptibility for a 3D spin system has been predicted from scaling
arguments by Chubukov {\it et al\/}. \cite{sigma-m} as $\chi(T)=aT^2$.
We calculate for the first time a universal factor in this
law. Employing a Quantum Monte Carlo cluster algorithm (QMC)
\cite{wiese} we then obtain a better estimate for the critical
coupling: $0.11<(J'/J)_c<0.12$. Next we calculate the temperature
dependence of the uniform susceptibility $\chi(T)$ for the whole
temperature range and various coupling ratios, shown in
Fig.~\ref{fig:chi}. 
Finally we show that the the susceptibility measurements of Hiroi and Takano 
\cite{Hiroi} can be fitted perfectly by the predicted form for a nearly
critical ordered system, thus resolving the apparent contradiction between the
susceptibility and magnetic resonance measurements.

A natural approach to the Hamiltonian (\ref{H}) from the side of
strong interladder coupling $J'\sim J$ is the renormalized spin-wave
theory of antiferromagnets \cite{Oguchi}.  Following a slightly
different procedure, we express the two spins per unit cell via two
types of boson operators $a_i$ and $b_i$ using the antiferromagnetic
Dyson-Maleev transformation.  Interaction terms with four bosons are
then treated in the mean-field approximation by introducing boson
averages: $m=\langle a_i^+a_i^{_{}}\rangle$, $\Delta_1=\langle
a_ib_i\rangle$, $\Delta_2=\langle a_ia_{i+z}\rangle$,
$\Delta_3=\langle a_ib_{i+x}\rangle$, which are determined by solving
self-consistent equations. The corresponding spin-wave spectrum
consists of two branches
\begin{eqnarray}
\omega_{\bf k}&=&\sqrt{A^2\!\!-\!(B_{\bf k}\!\pm\!|C_{\bf k}|)^2} \ ,  
\label{SW} \\
A &=& J(S\!-\!m\!+\!\Delta_1)\! +\! 2J(S\!-\!m\!+\!\Delta_2)\! +\! 
2J'(S\!-\!m\!+\!\Delta_3) , \nonumber\\
B_{\bf k}&=&2J(S\!-\!m\!+\!\Delta_2)\cos k_z \ , \nonumber\\
C_{\bf k}&=&J(S\!-\!m\!+\!\Delta_1)+
2J'(S\!-\!m\!+\!\Delta_3)(e^{ik_x}\!+e^{ik_y}) \ ,\nonumber 
\end{eqnarray}
each having zero-frequency mode at $k_z=0$ or $\pi$. At the isotropic 
point $J'=J$, our calculations predict for $S=1/2$ only a small 
reduction of the sublattice magnetization: $\langle S\rangle=0.40$. 
Quantum fluctuations destroy the magnetic order for the critical coupling 
$J'_c\approx0.05J$. (The result by the linear spin-wave theory
is an order of magnitude smaller.)
From general arguments we expect that the renormalized spin-wave 
theory overestimates the stability region of the ordered phase and, hence,
$J'_c\approx 0.05J$ presents a {\it lower} bound for
the exact critical value.

In the ordered phase the uniform magnetic susceptibility becomes
anisotropic with two components parallel and perpendicular to the
staggered moments. The parallel component $\chi^\parallel$ vanishes at
$T=0$. We calculate its low-temperature behavior in the framework of
the present approach by using
\begin{equation}
\chi^\parallel = \frac{1}{T} \sum_j \langle S_i^zS_j^z\rangle \ . 
\label{chisw1}
\end{equation}
In the limit $T\rightarrow0$ we find in agreement with Oguchi's 
results \cite{Oguchi}: $\chi^\parallel = T^2/6c_\parallel c_\perp^2$, 
where $c_\parallel$ and $c_\perp$ are the two spin-wave velocities 
determined from (\ref{SW}). The numerical coefficient in 
the square-law behavior of $\chi(T)$ increases by a factor 
of 20 between $J'=J$ and $J'_c$.

Describing correctly transverse oscillations in the ordered phase,
spin-wave theory fails, however, in the vicinity of $J'_c$ since at
the critical point excitation spectrum has the same triplet degeneracy
as in the disordered singlet phase for $J'<J'_c$.  To study the
order-disorder transition from the opposite side, we use the bond
operators formalism \cite{Sachdev}.  This method describes a single
spin ladder fairly well for strong enough rung coupling
\cite{Normand}. It has also been applied to a 3D array of ladders in
LaCuO$_{2.5}$ at $T=0$, but the result of Ref.~\cite{Normand} is
different from ours.

The two spins ($n=1,2$) belonging to the same ladder's rung
with the lattice index $i$ are expressed 
in terms of dimer states as
\begin{equation}
S_{n,i}^\alpha = \frac{(-1)^n}{2}(s_i^+t_{\alpha,i}+t_{\alpha,i}^+s_i)
 - \frac{i}{2} e^{\alpha\beta\gamma}t_{\beta,i}^+t_{\gamma,i} , 
\label{dimer}
\end{equation}
where $s_i$ and $t_{\alpha,i}$ are singlet and triplet boson operators
subject to the constraint 
$s_i^+s_i+\sum_\alpha t_{\alpha,i}^+t_{\alpha,i}^{} = 1$.
This relation is enforced by  
a chemical potential $\mu$. Also, a site independent 
condensate of singlets $\langle s_i\rangle  = \bar{s}$ is assumed. 
In the quadratic approximation we keep only the terms 
with two triplet operators. Diagonalizing the remaining Hamiltonian 
by the Bogoliubov transformation we obtain two self-consistent
equations $\langle\partial\hat{\cal H}_{\rm quad}/\partial\mu\rangle = 0$
and $\langle\partial\hat{\cal H}_{\rm quad}/\partial\bar{s}\rangle = 0$
on the parameters $\mu$ and $\bar s$. They can be reduced to a single
equation on the new parameter $d=2J\bar{s}^2/(J/4-\mu)$:
\begin{equation}
d = 5 - 6 \sum_{\bf k} \frac{1}{\sqrt{1+d\nu_{\bf k}}} 
\left(n_{\bf k} + \frac{1}{2}\right) , 
\label{selfc}
\end{equation}
where $\nu_{\bf k} = \cos k_z - J'/2J (\cos k_x + \cos k_y)$, 
magnon dispersion is $\omega_{\bf k}=(J/4-\mu)\sqrt{1+d\nu_{\bf k}}$,
and $n_{\bf k}$ is a Bose factor.
We first solve Eq.~(\ref{selfc}) at $T=0$. The gap becomes zero 
for $d=1/(1+J'/2J)$. Substituting this value into 
(\ref{selfc}) we find that the critical coupling corresponding
to vanishing gap and to the transition to the ordered phase is
$J'_c = 0.121J$. The mean-field theory should again overestimate
the stability region of the corresponding phase. Therefore,
we conclude that the above value is an {\it upper} bound
for the exact value of $J'_c$, which lies between 0.05 and 0.12.
We will find below from QMC that the exact
critical coupling is very close to the upper bound.
The spectrum of low-lying excitations in the disordered phase near 
$(0,0,\pi)$ has the form
$\omega_{\bf k} = c_\parallel\sqrt{k_z^2 + p^2 k_\perp^2 +m^2}$,
where $c_\parallel$ and $c_\perp=pc_\parallel$ are spin-wave
velocities parallel and perpendicular to ladders,
$c_\parallel\approx1.16J$ (at $J'=J'_c$), and $p=\sqrt{J'/2J}$.
The mass $m$ and the gap $\Delta=c_\parallel m$ behave like
$(J'_c-J')^{1/2}$ close to the critical point.

The isotropic susceptibility in the spin singlet state 
can be calculated by Eq.\ (\ref{chisw1}), which
after substitution of (\ref{dimer}) takes the form
\begin{equation}
\chi = \frac{1}{T} \sum_{\bf k} ( n_{\bf k}^2 + n_{\bf k}) \ ,
\label{chi}
\end{equation}
where summation is performed over one of the three magnon branches
only.

If the temperature is smaller than the gap, one can use the
zero-temperature spectrum. In this quantum disordered regime the
asymptotic behavior of the susceptibility found from Eq.\ (\ref{chi})
is
\begin{equation}
\chi(T) = \frac{\Delta^{3/2}T^{1/2}}{(2\pi)^{3/2}c_\parallel c_\perp^2} 
e^{-\Delta/T} \ ,
\label{chisb1}
\end{equation}
which differs by its prefactor from the analogous results for
magnetically disordered phases in 1D and 2D \cite{troyer}.

At $J'=J'_c$ the mass $m$ is generated by thermal fluctuations. It can
be found from the self-consistency equation at finite $T$. In contrast
to the 2D case \cite{Chubukov}, variation of  
the zero point fluctuation term in Eq.\ (\ref{selfc}) becomes 
logarithmically divergent on the upper limit, and is, therefore, lattice
dependent. Accordingly, $m$ is a linear function of $T$ with
logarithmically small prefactor computed by evaluating lattice sums:
\begin{equation}
\frac{c^2m^2}{T^2} = \frac{2\pi^2}{3 \ln(0.7J/T)}  \ .
\end{equation}
To calculate the universal behavior of the uniform susceptibility in
the quantum critical region $\Delta\ll T\ll J'$ we should neglect
logarithmically small mass and substitute the gapless dispersion into
Eq. (\ref{chi}). As a result, the universal form for the
susceptibility coincides with the result for $\chi^\parallel$ obtained
in the spin-wave theory:
\begin{equation}
\chi(T) = \frac{T^2}{6c_\parallel c_\perp^2} \ .
\label{chisb2}
\end{equation}
Notice that nonuniversal corrections to the prefactor in
the above expression have only logarithmic
smallness. 

Analogous calculations for the specific heat predict 
$C(T)=2\pi^2T^3/5c_\parallel c_\perp^2$ at the critical point.
The temperature dependence coincides again with the behavior
in the ordered phase. However, the prefactor is multiplied by 
$3/2$ according to the different number of gapless modes in
the two phases. Consequently, a crossover between these two
regimes should exist for a ``nearly critical'' ordered spin system. 

Critical behavior can be also studied using a sigma model description of
quantum antiferromagnets \cite{sigma-m}. Predictions
of that method have been compared with bond-operator
results for a 2D magnet in Ref.~\cite{Chubukov}.
By analogy we argue that the limit $N\rightarrow\infty$ 
of the $O(N)$ quantum nonlinear $\sigma$-model
in $3+1$ dimensions should give the same universal factor 
as in Eq.\ (\ref{chisb2}). 
This is quite natural
since both approaches use mean-field approximation.
Calculation of leading $1/N$ corrections to the mean-field
prediction remains an open question.

Using QMC we can obtain a better estimate for the critical coupling.
We have calculated the uniform susceptibility $\chi(T)$ for various
couplings on lattices up to $10\times 10$ ladders of length $40$ (8000
spins) and periodic boundary conditions at temperatures down to $\beta
J=24$.  The results are shown in Fig.~\ref{fig:chi}. We estimate the
critical coupling by varying the coupling ratio and looking for the
predicted $T^2$ behavior at criticality. Taking into account the shift
of the critical point due to finite size effects
\cite{finitesize} we estimate: $0.11 <(J'/J)_c < 0.12$, very close to
the bond-operator estimate.

Additionally we use self-consistent field boundary conditions 
\cite{scf} to probe the occurrence of N\'eel order 
and to estimate N\'eel temperatures. We find $T_N\approx0.38(3)J$ at
$J'/J=0.25$, $T_N\approx0.27(3)J$ at $J'/J=0.15$ and no indication for
order down to $T=J/16$ at $J'/J=0.1$. These results are consistent
with the above estimates and show that the N\'eel temperature of about
110K ($\approx J/10$) observed in the experiments is realized {\it
very close} to the critical point.

Next we want to discuss $\chi(T)$ for the whole temperature and
coupling range and compare with the experimental measurements. For all
couplings the Curie behavior at high temperatures changes over
into a broad maximum at temperatures of the order of $J$, caused by
local spin singlet formation on the individual ladders, just as in
uncoupled ladders \cite{troyer}. The single ladder then shows a
steep decrease with lowering the temperature, following an exponential
decay $\chi(T) \sim T^{-1/2}e^{-\Delta/T}$ \cite{troyer} with a gap of
about $0.5J$ at low temperatures \cite{review}.

A weak coupling between the ladders does not
destroy the spin gap. At high and intermediate temperatures we observe
the same behavior and a steep exponential decrease with a {\it pseudo
gap} similar to the gap of the single ladder. 
Only at temperatures
of the order of $J'$ a crossover to the 3D quantum disordered 
behavior Eq. (\ref{chisb1}), an exponential decay with the {\it actual
gap}, takes place.

When $\Delta$ becomes smaller than $J'$ upon approaching the $T=0$
transition point, the quantum critical region \cite{sigma-m} with its
$T^2$-law for the susceptibility appears between the quantum
disordered and decoupled ladders regimes.  Note, that existence of the
3D-type quantum critical behavior is restricted to quite low
temperatures $T<J'$. At $T>J'$, when interladder coupling can be
neglected, $\chi(T)$ still shows a remarkable similarity to the single
ladder.

In the ordered phase close to criticality we find the same pseudo gap
behavior, but the susceptibility goes to a small but nonzero value at
zero temperature. The crossover occurs at temperatures of the order of
the N\'eel temperature (compare $J'/J=0.15,0.2$ in
Fig. \ref{fig:chi}).

Let us now fit the susceptibility measurements on ${\rm
LaCuO}~{2.5}$. Hiroi and Takano have fitted them to an exponential
form plus a Curie contribution due to impurity spins, and thus
concluded a disordered ground state. But, as the magnetic resonance
measurements indicate an ordered ground state the correct low-T
behavior is
\begin{equation}
\label{eq:fit}
\chi(T)=C/(T-\Theta)+\chi_0+aT^2/J^3,
\end{equation}
 where $a\approx 0.33(3) {\rm e.m.u.\,mol}^{-1}$ estimated from QMC,
and $\chi_0$ is the sum of the temperature independent core
susceptibility, and van Vleck susceptibilty and the small
zero-temperature spin susceptibility.  The fit is excellent, as shown in
Fig. 3., with fitting parameters $C=1.8(1)\times 10^{-3}{\rm
e.m.u.\,mol}^{-1}$, $\Theta=-6.0(4){\rm K}$, $\chi_0=-6.2(2)\times
10^{-6}{\rm e.m.u.\,mol}^{-1}$, and $J=1340\pm 150{\rm K}$.

We see that the uniform susceptibility measured by Hiroi and Takano
\cite{Hiroi} is indeed compatible with a gapless ordered ground state
close to quantum criticality, as suggested by Normand and Rice
\cite{Normand}. We remark that due to the dominance of quantum
fluctuations in this nearly critical system no anomaly can be observed
at the N\'eel temperature.

Measurements of the total susceptibility suffer from the Curie
contribution of impurity spins at low temperatures, which make the
extraction of the asymptotic $T\rightarrow0$ behavior difficult. Thus
measurements which are not sensitive to impurities, such as NMR or
$\mu$SR are much better in distinguishing nearly critical ordered magnetic
materials from disordered ones.

We want to thank B. Normand and T.M. Rice for helpful discussions and
Z. Hiroi for providing us with data of their susceptibility
measurements. M.T. was supported by the Japan Society for the
Promotion of Science.

\begin{figure}
\caption{Cross section of the lattice structure of the model. The
ladders run perpendicular to the paper plane. Solid lines are the
rungs of the ladders with a coupling $J$. Dashed lines are the inter
ladder couplings $J'$. The dotted lines indicate the unit cell used.}
\label{fig:lattice}
\end{figure}

\begin{figure}
\caption{Uniform susceptibility calculated by QMC for some representative
ratios of the couplings.  Error bars were
omitted where the relative error was less than 1\%.
The inset is the same data in a double
logarithmic plot. The dotted line is added as a guide to the eye,
indicating the critical $T^2$ behavior.
clearer.}
\label{fig:chi}
\end{figure}

\begin{figure}
\caption{Fit of the susceptibility measurements by Hiroi and Takano
\protect{\cite{Hiroi}} to Eq. (\protect{\ref{eq:fit}}). The circles denote
the measurements, the solid line the fit, and the full circles the
measurements after subtraction of the Curie term.}
\label{fig:fitla}
\end{figure}

\end{document}